\documentclass{PoS}
\let\ifpdf\relax
\title{Recent  results from PHENIX on   double helicity asymmetry
   ($A_{LL}$) measurement at  $\sqrt{s}$ = 510 GeV}

\ShortTitle{Recent results from PHENIX on double helicity asymmetry ($A_{LL}$) measurements at $\sqrt{s}$ = 510 GeV}

\author{\speaker{H. Guragain} for the PHENIX Collaboration\\
        Georgia State University\\
        E-mail: \email{guragainhari@gmail.com}}

\abstract{One of  the major  objectives of the  RHIC  spin   program  at  BNL  is  to determine  the gluon helicity  contribution, $\Delta G$, to the spin of the proton. The PHENIX experiment probes  $\Delta G$ by measuring the double longitudinal spin asymmetry ($A_{LL}$) in the production of various
inclusive channels including  $\pi^{0}$ and $J/\psi$ mesons over a wide rapidity range. The $\pi^{0}$ is  reconstructed through  its di-photon decay  channel within the rapidity range of $\left|  \eta \right|$  $<$ 0.35  and azimuthal  angle of
$180\,^{\circ}$, while the $J/\psi$ is reconstructed via its dimuon decay channel within the rapidity range of 1.2 $<$ $\left|  \eta \right|$  $<$ 2.2. In 2013, the PHENIX experiment recorded an integrated luminosity of 150 pb$^{-1}$,  which is almost ten times the
total luminosity  recorded in 2009 at $\sqrt{s}$ =  200 GeV.
 The  increase  in  the   center  of  mass  energy  and  integrated
luminosity allows  covering the Bjorken x range down to $\sim$
0.01 for $\pi^{0}$  and $\sim$ 0.002 for $J/\psi$. Preliminary results for $A_{LL}$ in $\pi^{0}$ and $J/\psi$ production from the data collected in 2013 at  $\sqrt{s}$ =  510 GeV are presented, and their impact on $\Delta G$ is also discussed.}

\FullConference{The XXIII International Workshop on Deep Inelastic Scattering and Related Subjects\\
		April 27 - May 1, 2015\\
                Southern Methodist University\\
		Dallas, Texas 75275}

\begin{document}

\section{Introduction}
The inner spin structure of the nucleon can be characterized in terms of several key ingredients given by Manohar-Jaffe sum rule \cite{jaffe},\\
\begin{equation}
  S_{p} =\frac{1}{2} = \frac{1}{2}\Delta\Sigma  + \Delta G + L_{q} + L_{g},
\end{equation}
where $\Delta\Sigma$ is the combined quark and antiquark spin contribution, $L_{q}$ and $L_{g}$ are the quark and gluon orbital angular momentum contributions. $\Delta$G $\equiv$ $\int_{0}^{1}$dx$\Delta$g(x),  is the  gluon spin contribution to the total spin of proton and  $\Delta$g(x) is the gluon helicity distribution function of the nucleon which characterizes its internal structure.\\

High-energy scattering processes with polarized nucleons are used to access $\Delta\Sigma$ and  $\Delta$G. Polarized Deeply-Inelastic lepton-nucleon Scattering (DIS) data have shown that only about 25$\%$ of the total spin of proton is carried by the quark and antiquark spins ($\Delta\Sigma$ $\sim$ 25$\%$) \cite{florian,ball}. Unfortunately, the inclusive DIS measurements have small sensitivity to gluon contribution at present. Polarized proton-proton collisions at Relativistic Heavy Ion Collider (RHIC), Brookhaven National Laboratory (BNL) provide direct probes to the gluonic contribution ($\Delta$g) to the total spin of proton. The gluon helicity distribution can be accessed via a number of different channels, in particular jet or hadron production, as well as rarer probes such as direct photon and heavy flavor at high transverse momentum, $p_{T}$.

\section{Experimental Setup}
\begin{figure}[htp] 
\centering
\includegraphics[width=70mm,scale=0.6]{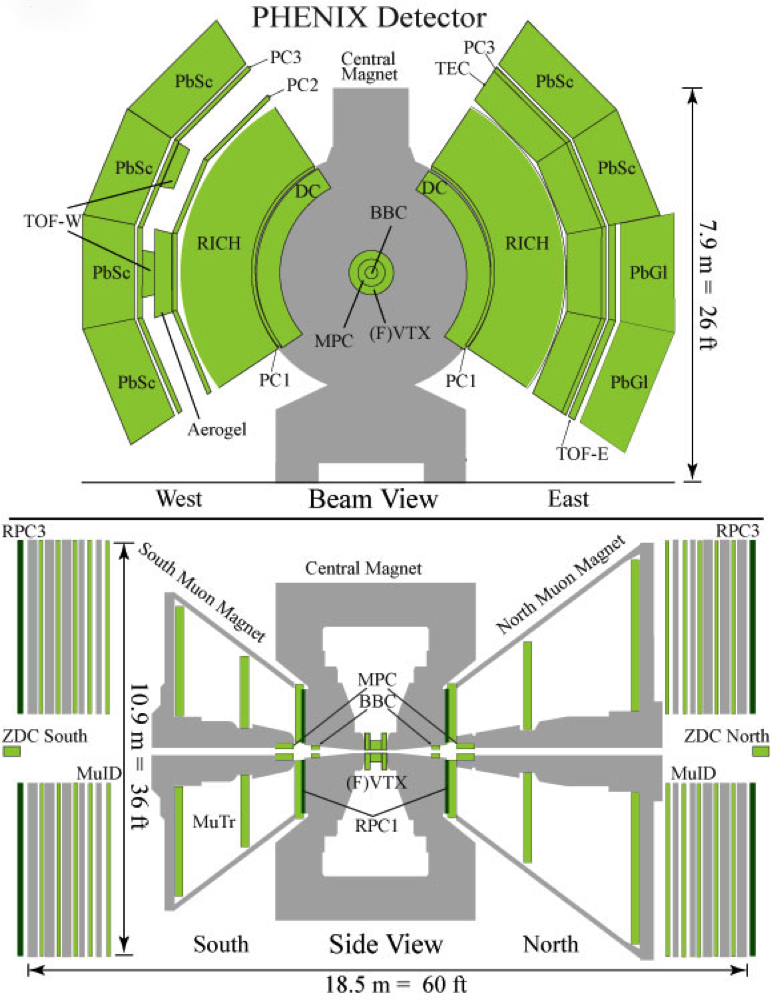}\\
\caption{A schematic beam- (top) and sideview (bottom) of the PHENIX
detector configuration in the 2013 data-taking period.}
\label{phnxdetc}
\end{figure}
The  highly segmented  PHENIX electromagnetic  calorimeter (EMCal)  is
used to  detect $\pi^{0}$ $\rightarrow$ $\gamma$  $\gamma$ decays. The
EMCal covers a  pseudorapidity range of $\mid$$\eta$ $\mid$$<$0.35
and azimuthal angle range of $\Delta$  $\phi$ = $\pi$. For
each of the two decay photons we require an energy deposition pattern consistent
with an  electromagnetic shower and  no charged track pointing  to the
location of the  deposited energy.  The details of the EMCal are described elsewhere \cite{emcal}.
The $J/\psi$ $\rightarrow$ $\mu^{+}$ $\mu^{-}$ is reconstructed in the muon arms. Each of the muon arms consists of the Muon Tracker (MuTr), immersed in a radial magnetic field of integrated bending power of 0.8 T$\cdot$m, and backed by the  Muon-Identifier (MuID). The muon arms cover the rapidities of -2.25 $\leq$ y $\leq$ -1.15 for the south arm and 1.15 $\leq$ y $\leq$ 2.44 for the north arm both with full azimuth \cite{overview}. Resistive plate chambers (RPC) detectors are used for the timing cuts in order to ensure the dimuon pair are produced  in the same crossing as the triggered crossing.\\

 The beam-beam counters (BBC) consist of two arms on opposite sides of the interaction-point (IP) along the beam axis at 3.1 < $|\eta|$ < 3.9. Each detector has 64 quartz crystal Cerenkov radiators attached to photomultiplier tubes (PMTs). The BBC is used for luminosity measurement and triggering. The second luminosity monitor, the zero degree calorimeter (ZDC), consists of two arms located $|z|$ = 18 m from the Interaction Point (IP) along the beam axis, covering  $|\eta|$ > 6. Each arm is composed of three sections of tungsten-scintillator sandwich calorimetry with a total of 5 nuclear interaction lengths.

\section{Results}
The double helicity asymmetry can be written as
\begin{equation}
  A_{LL} = \frac{\Delta\sigma}{\sigma} = \frac{\sigma_{++} - \sigma_{+-}}{\sigma_{++} + \sigma_{+-}}
\end{equation}
where $\sigma_{++}$($\sigma_{+-}$) is the cross section for same (opposite) helicity collisions. In terms of yield (N) and luminosity (L), this formula can be rewritten as:\\
\begin{equation}
A_{LL} = \frac{1}{P_{B}P_{Y}}\frac{N^{++}-RN^{+-}}{N^{++}+RN^{+-}},\enspace  R\equiv \frac{L^{++}}{L^{+-}}
\end{equation}
where $P_{B(Y)}$ is the polarization of the blue (yellow) beam, R is the relative luminosity and $N^{++}(N^{+-})$ are the contributions from the same-sign (opposite sign) helicity collisions.\\
\vspace{10 mm}
The statistical uncertainty of $A_{LL}^{\pi^{0}}$ is calculated using the following formula:\\
\begin{equation}
\sigma_{A_{LL}} = \frac{1}{P_{B}P_{Y}}\frac{2RN^{++}N^{+-}}{\left(N^{++}+RN^{+-}\right)^{2}}\sqrt{\left(\frac{\sigma_{N^{++}}}{N^{++}}\right)^{2}+\left(\frac{\sigma_{N^{+-}}}{N^{+-}}\right)^{2}+\left(\frac{\sigma_{R}}{R}\right)^{2}}
\end{equation}
where $\sigma_{N^{++}}$ ($\sigma_{N^{+-}}$) are the statistical uncertainty in same (opposite) helicity collisions.
\begin{figure}[htp!] 
\centering                      
\includegraphics[height=7.0cm,width=9cm]{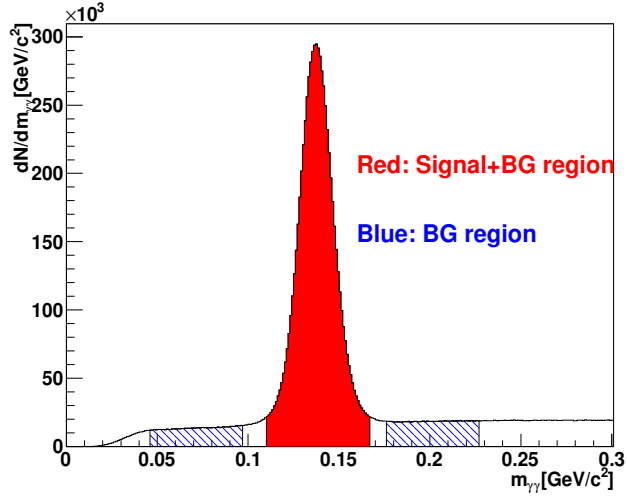}
\caption{Invariant mass spectrum of $\pi^{0}$. The signal region is represented by red region (0.112  $\leq$ $M_{\gamma\gamma}$ $\leq$ 0.162 GeV/c$^{2}$) while the background region is represented by blue region (0.047  $\leq$ $M_{\gamma\gamma}$ $\leq$ 0.097 GeV/c$^{2}$ and 0.177 $\leq$ $M_{\gamma\gamma}$ $\leq$ 0.227 GeV/c$^{2}$)}
\label{massSpectrum}
\end{figure}

Figure \ref{massSpectrum} shows  the invariant mass spectra from all photon pairs. The $\pi^{0}$ mass-peak region  includes both signal and background. The asymmetry measured in this region, $A_{LL}^{S+BG}$  is a mix of both the signal asymmetry, $A_{LL}^{S}$, and the asymmetry in the background, $A_{LL}^{BG}$.  The relationship between these three asymmetries in the mass peak region can be written as\\
\begin{equation}\label{eq:ldsa}
A_{LL}^{\pi^{0}} = \frac{A_{LL}^{\pi^{0}+BG}-r A_{LL}^{BG}}{1 - r}
\end{equation}
\begin{equation}\label{eq:sldsa}
  \sigma_{A_{LL}}^{\pi^{0}} = \frac{\sqrt{{\sigma}^{2}_{A_{LL}^{\pi^{0}+BG}}+r^{2} {\sigma}^{2}_{A_{LL}^{BG}}}}{1 - r}
\end{equation}

where $\textit{r}$ is the background fraction in the signal region (0.112 $<$ $M_{\gamma\gamma}$ $<$ 0.162 GeV/c$^{2}$), which corresponds to $\sim$ 3$\sigma$ about  the  mean  of  the  mass  peak.   The background ratio, $\textit{r}$, is the ratio of the background in the signal region extracted from fitting the mass spectrum with Gaussian plus a second order polynomial and the integration of invariant mass spectra in the signal region. $A_{LL}^{BG}$ is calculated in the background regions (0.047  $\leq$ $M_{\gamma\gamma}$ $\leq$ 0.097 GeV/c$^{2}$ and 0.177 $\leq$ $M_{\gamma\gamma}$ $\leq$ 0.227 GeV/c$^{2}$).

\begin{figure}[htp!] 
    \centering \includegraphics[height=8.8cm,width=11cm]{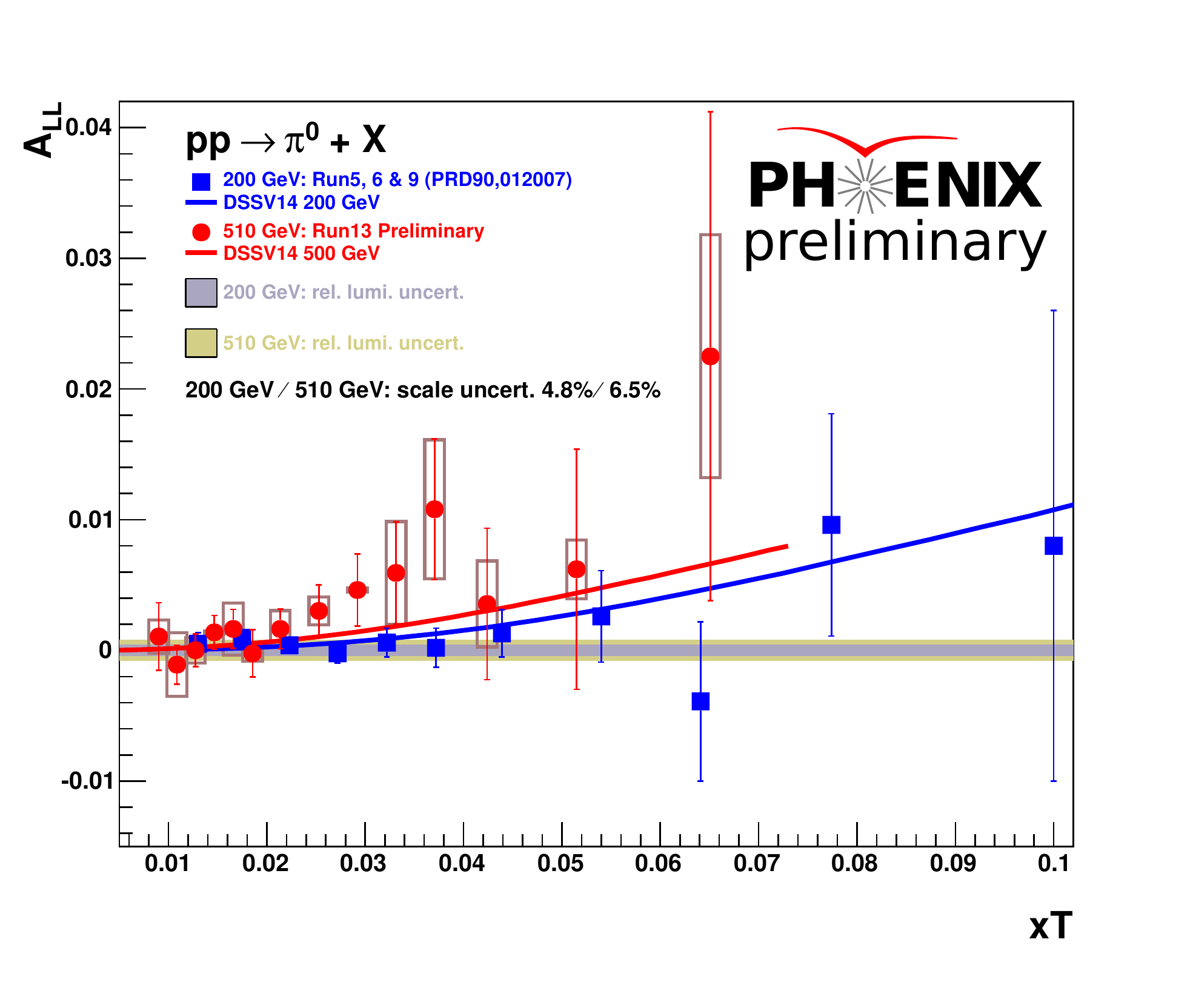}\\
    \caption{$A_{LL}$ for $\pi^{0}$ as a function  of $x_{T}$.}
    \label{lda}
\end{figure}

\begin{figure}[htp!] 
\centering                      
\includegraphics[height=8.8cm,width=11cm]{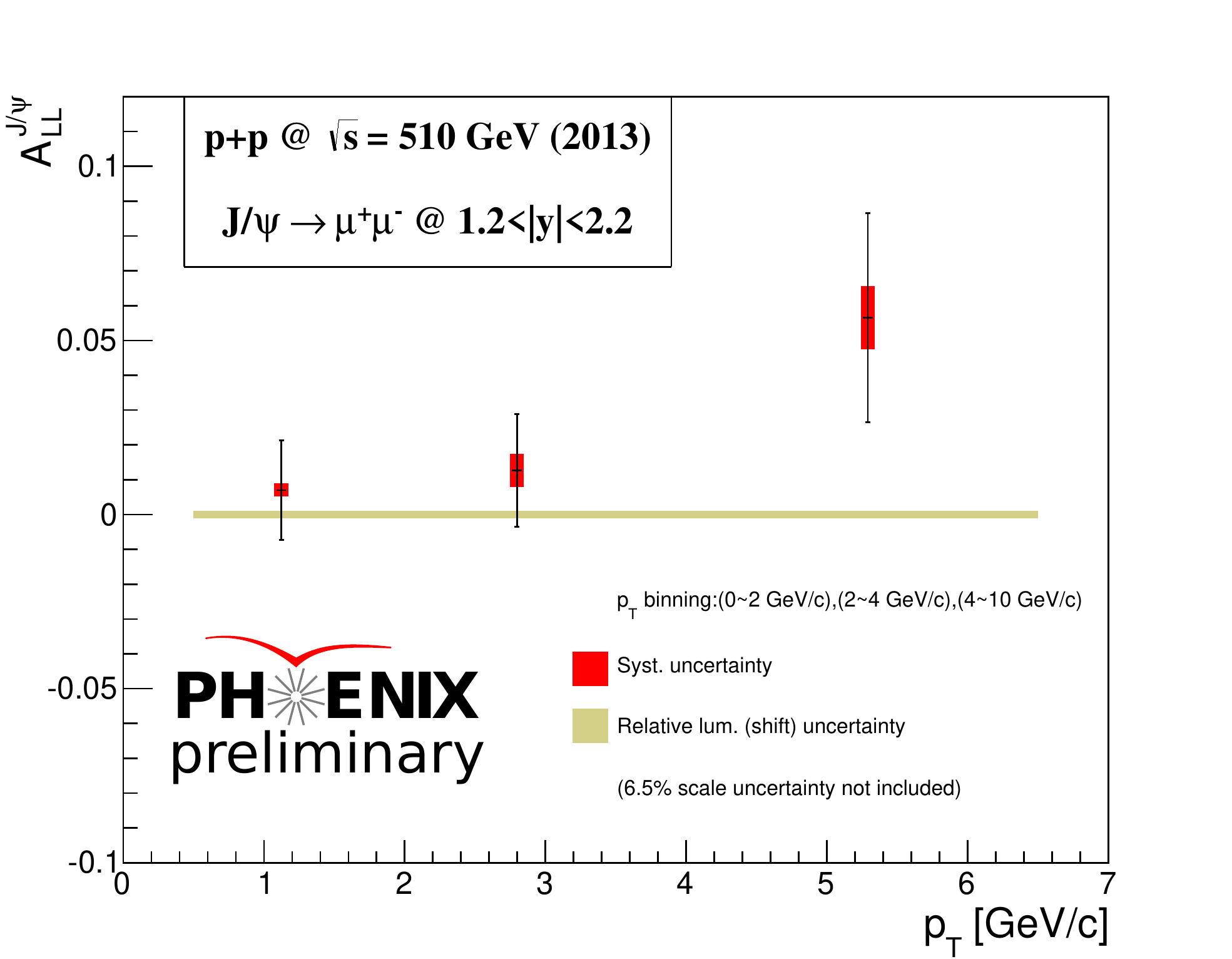}
\caption{$A_{LL}$ for  $J/\psi$ as a function  of $p_{T}$.}
\label{jpsi}
\end{figure}

The same formula is used to extract the $A_{LL}$ and corresponding statistical errors for $J/\psi$. However, the crystal ball function is used to extract the $J/\psi$ signal and gaussian process regression (GPR) is used for the background fit.
The $\pi^{0}$ $A_{LL}$ is calculated within the $p_{T}$ range  2$\leq$ $p_{T}$ $\leq$ 20 GeV/c while  the $J/\psi$  $A_{LL}$ is calculated within the $p_{T}$ range 1$\leq$ $p_{T}$ $\leq$ 6 GeV/c.\\
\begin{figure}[htp!]
    \centering
    \includegraphics[height=8.8cm,width=11cm]{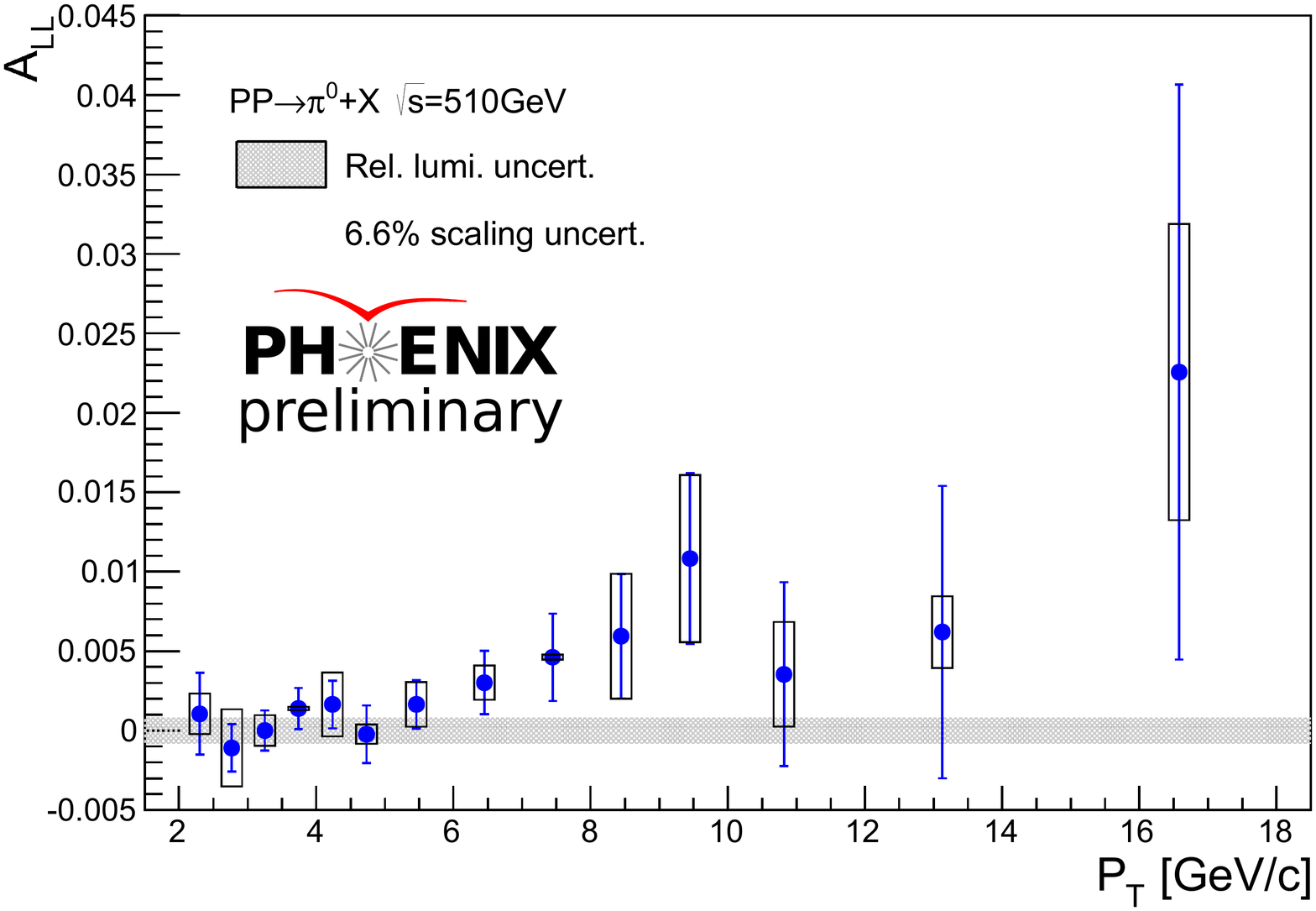}\\
    \caption{$A_{LL}$ for $\pi^{0}$ as a function  of $p_{T}$.}
    \label{fpt}
\end{figure}
Figure \ref{lda} shows the measured $A_{LL}^{\pi^{0}}$ as the function of $x_{T}$ = ($\frac{2P_{T}}{\sqrt{s}}$). The results are compared with the DSSV calculations \cite{florian}. The gray bands show the associated systematic uncertainties. While the data are consistent with the DSSV calculation within uncertainties at low pT, significant non-zero asymmetry is observed which  indicates a non-zero gluon contribution to the spin of the proton in the accessed kinematic range.  $A_{LL}^{\pi^{0}}$ at $\sqrt{s}$ = 200 GeV is also shown in Fig. \ref{lda}. Figure \ref{fpt} shows the $A_{LL}^{\pi^{0}}$ as the function of $p_{T}$. Figure \ref{jpsi} shows the measured  $A_{LL}^{J/\psi}$ as the function of $p_{T}$.\\

\section{Summary}
We presented recent PHENIX measurements of $A_{LL}$ in $\pi^{0}$ and  $J/\psi$  production in longitudinally polarized p + p collisions at $\sqrt{s}$ = 510 GeV based on the data collected in the year 2013. A significant non-zero asymmetry is observed indicating a non-zero gluon contribution to the spin of the proton in the accessed kinematic range.

\end{document}